\documentclass[reprint,aps,prb,superscriptaddress,floatfix]{revtex4-1}

\usepackage{amssymb,amsmath,amsfonts,mathrsfs,bm}
\usepackage{graphicx}
\usepackage{color}
\usepackage{hyperref}
\usepackage{listings}
\usepackage{ulem}
\usepackage{newtxmath}
\hypersetup{
    unicode=false,          
    pdftoolbar=true,        
    pdfmenubar=true,        
    pdffitwindow=false,     
    pdfstartview={FitH},    
    colorlinks=true,       
    linkcolor=blue,          
    citecolor=blue,        
    urlcolor=blue,
}

\begin{document}

\title{Perdew-Zunger self-interaction correction: How wrong for uniform densities and large-$Z$ atoms?}
\author{Biswajit Santra}
\email{biswajit.santra@temple.edu}
\email{bishalya@gmail.com}
\affiliation{Department of Physics, Temple University, Philadelphia, PA 19122}
\author{John P. Perdew}
\affiliation{Department of Physics, Temple University, Philadelphia, PA 19122}
\affiliation{Department of Chemistry, Temple University, Philadelphia, PA 19122}
\date{\today}

\begin{abstract}
Semi-local density functionals for the exchange-correlation energy of a many-electron system cannot be exact for all one-electron densities. In 1981, Perdew and Zunger (PZ) subtracted the fully-nonlocal self-interaction error orbital-by-orbital, making the corrected functional exact for all collections of separated one-electron densities, and making no correction to the exact functional. Although the PZ self-interaction correction (SIC) eliminates many errors of semi-local functionals, it is often worse for equilibrium properties of $sp$-bonded molecules and solids. Non-empirical semi-local functionals are usually designed to be exact for electron gases of uniform density, and thus also make 0\% error for neutral atoms in the limit of large atomic number $Z$, but PZ SIC is not so designed. For localized SIC orbitals, we show analytically that the LSDA-SIC correlation energy per electron of the uniform gas in the high-density limit makes an error of -50$\%$ in the spin-unpolarized case, and -100\% in the fully-spin-polarized case. Then we extrapolate from the Ne, Ar, Kr, and Xe atoms to estimate the relative errors of the PZ SIC exchange-correlation energies (with localized SIC orbitals) in the limit of large atomic number: about +5.5\% for the local spin density approximation (LSDA-SIC), and about -3.5\% for nonempirical generalized gradient (PBE-SIC) and meta-generalized gradient (SCAN-SIC) approximations. The SIC errors are considerably larger than those that have been estimated for LSDA-SIC by approximating the localized SIC orbitals for the uniform gas, and may explain the errors of PZ SIC for equilibrium properties, opening the door to a generalized SIC that is more widely accurate.

\end{abstract}

\maketitle

\section{\label{sec:one}Introduction}
A realistic description of bonding between atoms in molecules and solids within Kohn-Sham (KS) ground-state density functional theory \cite{Kohn1965,Jones2015} requires a realistic approximation to the exact spin-density functional for the exchange-correlation energy \cite{Jones2015}. Local or semi-local functionals such as the local spin density approximation (LSDA), the generalized gradient approximation (GGA), and the meta-GGA are widely employed. These functionals are single integrals over position space of energy densities that are functions of ingredients readily available at each position from self-consistent solution of the KS or generalized KS one-electron Schr\"odinger equation. Standard ingredients include the spin-resolved electron densities (LSDA), and sometimes also their gradients (GGA) and even orbital kinetic energy densities (meta-GGA). Semi-local functionals are computationally efficient and often usefully accurate, and lend themselves to non-empirical construction by satisfaction of many exact constraints and appropriate norms \cite{Sun2015}.  The earliest appropriate norms were the uniform electron gases \cite{Kohn1965,Perdew1992}, for which nearly all practical functionals are exact by construction.

Perhaps the simplest density is that for one isolated electron (e.g., $n_\uparrow=n=|\phi|^2,n_\downarrow=0$), for which the exact exchange-correlation energy must exactly cancel the spurious Hartree electrostatic self-interaction. But, since the Hartree energy 
\begin{equation}
\label{eq:1}
U[n] = \frac{1}{2}\int d^3r \, n(\bm{r}) \int d^3r^\prime \, n(\bm{r}^\prime) / |\bm{r}-\bm{r}^\prime|
\end{equation} 
is a fully-nonlocal functional of the density, no local or semi-local approximation can be exact for all possible one-electron densities. In 1981, Perdew and Zunger (PZ) \cite{Perdew1981} proposed to subtract the spurious self-interaction from each occupied orbital,
\begin{align}
\label{eq:2}
E_{xc}^{approx-SIC} & = E_{xc}^{approx}[n_\uparrow,n_\downarrow] \, - \nonumber \\ & \sum_{i\sigma}^{occup}\{U[n_{i\sigma}]+E_{xc}^{approx}[n_{i\sigma},0]\} \,,
\end{align}
where $n_{i\sigma}=f_{i\sigma}|\phi_{i\sigma}|^2$ is the $i$-th occupied ($f_{i\sigma}=1$) orbital density of spin $\sigma$, and thus a one-electron density. Eq. (\ref{eq:2}) is exact for any collection of ground- or excited-state one-electron densities. PZ self-interaction correction (SIC), applied to the exact functional, makes no change in it.

PZ also realized that the right orbitals for SIC cannot be the often-delocalized Kohn-Sham orbitals, but must be localized to make a properly size-extensive theory in which the energy of a periodic array of $N$ separated atoms always becomes $N$ times the energy of one atom in the limit of large separation. Lin and collaborators \cite{Pederson1984} found unitary transformations in the occupied-orbital space that minimize the PZ SIC energy. More recently, Pederson \textit{et al.} \cite{Pederson2014} have found a way to limit this transformation to Fermi-L\"owdin orbitals (FLO’s) that are guaranteed to be localized, because each has a Fermi orbital descriptor $\bm{r}_{i\sigma}$ or position where
\begin{equation}
\label{eq:3}
|\phi_{i\sigma}(\bm{r}_{i\sigma})|^2=n_{\sigma}(\bm{r}_{i\sigma}) \,\,.
\end{equation}
PZ SIC leads to an almost linear (and thus almost exact \cite{Perdew1982}) variation of the energy of an open system with non-integer average electron number, between adjacent integer electron numbers, and thus often avoids spurious fractional charges on dissociated atoms \cite{Ruzsinszky2006}. And it corrects many other qualitative failures of the semi-local functionals, yielding a correct exchange-correlation potential at the edges of a system, bound negative atomic ions, raised and improved energy barriers to chemical reactions \cite{Patchkovskii2002}, etc.

For equilibrium properties like atomization energies and equilibrium bond lengths, the non-empirical semi-local functionals have improved in accuracy from LSDA \cite{Kohn1965,Perdew1992} to GGA (e.g., Ref. \onlinecite{Perdew1996}) and meta-GGA (e.g., Ref. \onlinecite{Sun2015}). But PZ SIC applied to GGA \cite{Vydrov2004,Vydrov2006} and meta-GGA \cite{Shahi2019} is actually less accurate than GGA or meta-GGA, respectively, for these properties. Part of the explanation for this lies in the noded character of overlapped orthogonal localized real orbitals. Using complex SIC orbitals leads to a lower total energy, and reduces but does not eliminate the errors in the equilibrium properties \cite{Klupfel2011}. We have recently argued \cite{Shahi2019} that lobed one-electron densities (including the familiar stretched one-electron densities and the less familiar noded ones) are problematic for semi-local approximation. And some of the troublesome lobed character survives after the nodes are removed.

Another part of the explanation could be that PZ SIC loses the exact uniform-density limit that the non-empirical semi-local functionals all respect. We now know that, in the limit of large atomic number $Z$, most of the electron density in a neutral atom is slowly-varying on the scale of the local Fermi wavelength \cite{Perdew2006}, and that the local density approximation, based only upon the uniform electron gas, becomes relatively exact for the exchange energy and even for the correlation energy (according to numerical evidence) \cite{Burke2016}. The exact large-$Z$ asymptotic expansions for the exchange \cite{Schwinger1981,Elliott2009} and correlation \cite{Burke2016,Cancio2018} energies of neutral atoms are:
\begin{equation}
\label{eq:4}
E_x=-A_x \, Z^{\frac{5}{3}} + B_x \, Z + C_x \, Z^{\frac{2}{3}} + \cdots \,\,,
\end{equation}
\begin{equation}
\label{eq:5}
E_c=-A_c \, Z \, ln(Z) + B_c \, Z + \cdots \,\,.
\end{equation}
Clearly the ratio $E_c/E_x$ tends to zero as $Z$ tends to infinity. LSDA \cite{Kohn1965,Perdew1992}, the PBE GGA \cite{Perdew1996}, and the SCAN (strongly-constrained and appropriately-normed) meta-GGA \cite{Sun2015} have similar expansions, and all of those semi-local approximations are exact to leading order. PBE and SCAN, but not LSDA, are also accurate for the next-order terms in $Z$ in Eqs. (\ref{eq:4}) and (\ref{eq:5}), as shown in Table \ref{tab:1}. (While SCAN was fitted to the exact large-$Z$ limits, the values of its coefficients for Eqs. (\ref{eq:4}) and (\ref{eq:5}) are unpublished.)


\begin{table}[h!]
\caption{\label{tab:1}Known coefficients of the large-$Z$ asymptotic expansions for the exchange energy of Eq. (\ref{eq:4}) and the correlation energy of Eq. (\ref{eq:5}) for neutral atoms.}
\begin{ruledtabular}
\begin{tabular}{ccccc}
Functional & $A_x$ & $B_x$ & $A_c$ & $B_c$ \\ 
\hline
Exact  & 0.2209\cite{Cancio2018}   &    -0.224\cite{Burke2016}  &   0.02073\cite{Cancio2018}    &    0.0372\cite{Burke2016}  \\
LSDA   & 0.2209\cite{Cancio2018}   &    -0.0\cite{Burke2016}    &   0.02073\cite{Cancio2018}    &   -0.00451\cite{Cancio2018} \\
PBE    & 0.2209\cite{Cancio2018}   &    -0.215\cite{Elliott2009}&   0.02073\cite{Cancio2018}    &    0.03936\cite{Burke2016}
\end{tabular}
\end{ruledtabular}
\label{tab:1}
\end{table}

Before the relevance of the uniform gas to real systems was fully understood, it was hard to determine the errors of PZ SIC for the uniform electron gas, because there was no reliable construction of the localized SIC orbitals. Relative errors in the PZ LSDA-SIC negative exchange energy of about -2.5\% \cite{Norman1983} or about +0.8\% or more \cite{Pederson1989} were estimated. But we can readily construct the energy-minimizing real Fermi-L\"owdin orbitals for the atoms Ne, Ar, Kr, and Xe, then extrapolate the exchange energy to large $Z$ using the known asymptotic expansions of Eqs. (\ref{eq:4}) and (\ref{eq:5}). The resulting relative errors in the negative exchange energy turn out to about +5.5\% for PZ LSDA-SIC and about -3.5\% for PZ PBE-SIC and PZ SCAN-SIC, and thus not negligible. Since $E_x$ and $E_c$ are negative, a positive percentage error means that an approximation is too negative, and a negative percentage error means that the approximation is not negative enough.

In the high-density limit, exchange completely dominates over correlation. Since the PZ LSDA-SIC exchange energy per electron of a uniform electron gas is too low with real SIC orbitals, minimizing the total SIC energy over  complex orbitals can only make it lower and less accurate. A consequence of our work is that, in a high-density uniform electron gas, a real unitary transformation that minimizes the total energy will yield localized SIC orbitals in PZ LSDA-SIC, but (unless constrained as in the Fermi-L\"owdin-orbital approach\cite{Pederson2014}) delocalized SIC orbitals in PZ PBE-SIC and PZ SCAN-SIC. Although delocalized or Kohn-Sham SIC orbitals would make the PZ PBE-SIC and PZ SCAN-SIC exact for uniform electron gases, they are ruled out by our requirement that SIC be size extensive.

While the exact exchange energy of Kohn-Sham theory requires a multiplicative exchange potential $V_{x,\sigma}(r)$, it is for practical purposes equivalent to the Hartree-Fock exchange energy for the same electron spin densities. The difference is only about 0.03\% for the Ar atom\cite{Gorling1995}, and tends to 0\% as $Z\rightarrow \infty$.
The exact exchange energy and all reasonable approximations to it (including Eq. (\ref{eq:2}) with $xc$ replaced everywhere by $x$) are homogeneous functionals that obey a uniform-density-scaling equality \cite{Levy1985}. Thus, while the large-$Z$ limit for neutral atoms corresponds to high electron densities, the relative (\%) error of a reasonable approximation to the exchange energy functional is the same for all uniform electron densities.

\begin{table*}[htb]
\caption{\label{tab:2} Exchange energy for neutral rare-gas atoms (Ne, Ar, Kr, Xe) with atomic number $Z$. The energies are from reference Hartree-Fock (taken from Ref. \onlinecite{Becke1988}) and six non-empirical DFT functionals computed in this work. Self consistent electron density is used for all DFT functionals. Energies are given in Hartree unit.}
\begin{ruledtabular}
\begin{tabular}{cccccccc}
$Z$ & $E_x^{exact}$ & $E_x^{LSDA}$ & $E_x^{PBE}$ & $E_x^{SCAN}$ & $E_x^{LSDA-SIC}$ & $E_x^{PBE-SIC}$ & $E_x^{SCAN-SIC}$ \\ 
\hline
10 & -12.108  & -10.9668  &  -12.0269 &   -12.1413 &  -12.4636 &  -12.0385 &  -11.9934 \\
18 & -30.188  & -27.8122  &  -29.9697 &   -30.2563 &  -31.1554 &  -29.8464 &  -29.7817 \\
36 & -93.890  & -88.5356  &  -93.3685 &   -94.0488 &  -97.6783 &  -92.3723 &  -92.1973 \\
54 & -179.200 & -170.5124 & -178.2248 &  -179.2896 & -186.5765 & -175.5691 & -175.2490 \\
\end{tabular}
\end{ruledtabular}
\end{table*}

\begin{table*}[htb]
\caption{\label{tab:3} Correlation energy for neutral rare-gas atoms (Ne, Ar, Kr, Xe) with atomic number $Z$. The numbers are from reference calculation (taken from Ref. \onlinecite{Burke2016}) and six non-empirical DFT functionals computed in this work. Self consistent electron density is used for all DFT functionals. Energies are given in Hartree unit.}
\begin{ruledtabular}
\begin{tabular}{cccccccc}
$Z$ & $E_c^{exact}$ & $E_c^{LSDA}$ & $E_c^{PBE}$ & $E_c^{SCAN}$ & $E_c^{LSDA-SIC}$ & $E_c^{PBE-SIC}$ & $E_c^{SCAN-SIC}$ \\ 
\hline
10 & -0.3910  & -0.7398  & -0.3469  & -0.3444 & -0.4108 & -0.2834 & -0.3446 \\
18 & -0.7254  & -1.4232  & -0.7042  & -0.6916 & -0.7952 & -0.5718 & -0.6911 \\
36 & -1.8504  & -3.2683  & -1.7632  & -1.7591 & -1.8579 & -1.4797 & -1.7584 \\
54 & -3.0002  & -5.1782  & -2.9159  & -2.9059 & -2.9586 & -2.4692 & -2.9047 \\
\end{tabular}
\end{ruledtabular}
\end{table*}

Since SCAN is self-correlation-free, it requires and gets no SIC for correlation, but the LSDA and PBE correlation functionals do. The leading term for the correlation energy in the high-density limit, with or without SIC, can be found exactly by analytic derivation, and will thus be presented before and used in the extrapolation for the exchange-correlation energy. The PZ SIC relative errors for the correlation energy in the high-density limit are -50\% to -100\%, for the spin-unpolarized and fully-polarized cases respectively. The special property of the logarithm, $ln(AB) = ln(A)+ln(B)$, makes it possible to find these relative errors without constructing the localized SIC orbitals for the uniform electron gas. 

Understanding the errors that PZ SIC makes for uniform densities and for large-$Z$ atoms is a necessary first step toward the development of a generalized SIC \cite{Perdew2019} that would be more widely accurate than PZ SIC. Such a correction might modify the semi-local part of the PZ SIC of Eq. (\ref{eq:2}) in regions where the localized SIC orbitals overlap, as in Ref. \onlinecite{Vydrov2006} but without modifying the fully non-local part. 
             
Our calculations for atoms are made with the FLOSIC \cite{flosic0.1} molecular code, which in turn is based upon and uses the default all-electron Gaussian basis set of the NRLMOL code \cite{Porezag1999}. The basis sets allow us to treat Ne, Ar, Kr, and Xe, but not Rn (which has $f$ electrons). The integration mesh for SCAN is about three times bigger than the default mesh. The integration mesh in the FLOSIC code is variational \cite{Pederson1990} and for all calculations the maximal error margin for the integration mesh is set to $10^{-7}$. The descriptors or positions of the FLOs are optimized using LSDA-SIC until the maximum component of the forces\cite{Pederson2015} are lower than $10^{-3}$ Hartree/Bohr. The LSDA-SIC optimized descriptors are used to self-consistently minimize the FLOs and the corresponding PZ SIC total energy \cite{Yang2017} with a threshold of $10^{-6}$ Hartree for all PZ SIC functionals employed here. The exchange and correlation energies obtained from these calculations are given in Tables \ref{tab:2} and \ref{tab:3}.

\section{\label{sec:2}Correlation energy at high density: Analytic derivation}

The exact correlation energy per electron for a uniform electron gas in the high-density limit is given by the random phase approximation\cite{Gell-Mann1957,Misawa1965}:
\begin{equation}
\label{eq:6}
e_c^{exact}(n,\zeta)=-\alpha_c(\zeta) \, ln(n)
\end{equation}
where $n=n_\uparrow+n_\downarrow$ is the total density and $\zeta=(n_\uparrow-n_\downarrow)/n$ is the relative spin polarization (0 for the unpolarized case, and $\pm 1$ for the fully spin-polarized case). Note that\cite{Misawa1965} $\alpha_c(1)=\frac{1}{2}\alpha_c(0)$. We are interested in the unpolarized case, but will need the fully-polarized case to make the SIC.

In PZ SIC the localized SIC orbital density is
\begin{equation}
\label{eq:7}
n_{i\sigma}({\bf r})= n \, g_{i\sigma}(n^{1/3}\bf r) \,\,.
\end{equation}
Eq. (\ref{eq:7}) has the proper uniform-density scaling\cite{Levy1985}, can integrate to 1 for all $n$, and can satisfy Eq. (\ref{eq:3}) for all $n$. The shape of this orbital density is unknown, but will not contribute to the correlation energy per electron of order $ln(n)$. Thus the relative error in the high-density limit will be the same whether the SIC orbitals are real or complex. For LSDA, which are exact for all uniform densities, the PZ SIC correlation energy per electron of the spin-unpolarized uniform electron gas in the high-density limit is
\begin{align}
\label{eq:8}
e_c^{LSDA-SIC}(n,0)&=-\alpha_c(0) \, ln(n) \, - \nonumber \\ & \frac{1}{N} \sum_{i\sigma} \int d^3r \, n_{i\sigma}({\bf r}) [-\alpha_c(1) \, ln( n_{i\sigma}({\bf r}))] \,\, .
\end{align}
Using Eq. (\ref{eq:7}) and applying the special property of the logarithm $ln(n \, g_{i\sigma}(n^{1/3}{\bf r}))=ln(n)+ln(g_{i\sigma}(n^{1/3}{\bf r}))$ to Eq. (\ref{eq:8}) we obtain:
\begin{widetext}
\begin{align}
\label{eq:9}
e_c^{LSDA-SIC}(n,0)&=-\alpha_c(0) \, ln(n) \, + 
\alpha_c(1) \, ln(n) \, \frac{1}{N} \sum_{i\sigma} \int d^3r \, n_{i\sigma}({\bf r}) \, +
\alpha_c(1) \, \frac{1}{N} \sum_{i\sigma} \int d^3(n^{1/3}{\bf r}) \, g_{i\sigma}(n^{1/3}{\bf r}) \, ln (g_{i\sigma}(n^{1/3}{\bf r})) \,\,.
\end{align}
\end{widetext}
Neglecting the lower-order term (a constant independent of $n$) and using $\alpha_c(1)=\frac{1}{2}\alpha_c(0)$, we find
\begin{equation}
\label{eq:10}
e_c^{LSDA-SIC}(n,0)=-\frac{1}{2}\alpha_c(0) \, ln (n) = \frac{1}{2} e_c^{exact}(n,0) \,\,,
\end{equation}
for a -50\% error. Following similar reasoning, the PZ LSDA-SIC correlation energy per electron for fully spin-polarized uniform electron gas in the high-density limit is 0 times $ln(n)$, for a -100\% error.

In the large-$Z$ limit the exact correlation energy of the (spin-unpolarized) atom is $E_c^{exact}=-A_c \, Z\, ln (Z)$, where $A_c=0.02073$\cite{Cancio2018}. Thus in the large-$Z$ limit the LSDA-SIC correlation energy is
\begin{equation}
\label{eq:11}
E_c^{LSDA-SIC}=-\frac{A_c}{2} Z \, ln (Z) \,,
\end{equation}
for a -50\% error.

For the PBE-SIC correlation energy, however, the relative error in the large-$Z$ limit is 0\%. This follows from Eq. (9) of Ref. \onlinecite{Perdew1996}, and from the fact that the reduced density gradient $s$ for the SIC orbital density does not approach 0 in the large-$Z$ limit. As a result, the self-interaction correction to the PBE correlation energy per electron scales as $Z^0$, and not as $ln(Z)$, when $Z\rightarrow \infty$.

\begin{figure}[htb]
\centering
\includegraphics[width=1.0\linewidth]{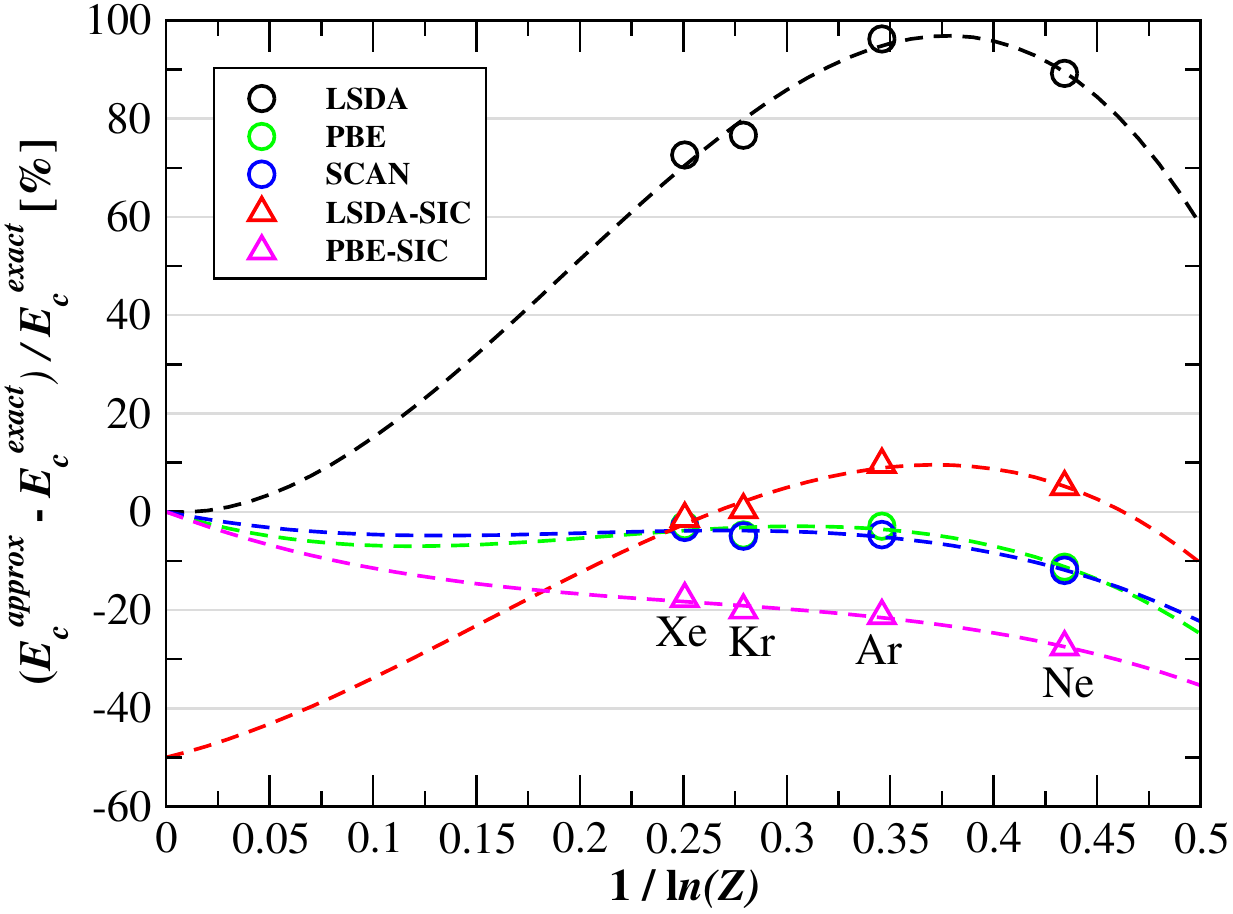}
\caption{Relative (\%) errors of five nonempirical functionals for the correlation energies of neutral atoms with atomic number $Z$. (Note that for correlation SCAN-SIC=SCAN.) The points shown are for (from left to right) Xe ($Z=54$), Kr ($Z=36$), Ar ($Z=18$), and Ne ($Z=10$). The smooth curves are based upon least-squares fits of the parameters $a$, $b$, and $c$ in the asymptotic formula of Eq. (\ref{eq:12}). The $1/ln(Z)\rightarrow 0$ limit gives the relative error of the correlation energy for the high-density spin-unpolarized uniform electron gas: $\Delta=0$ for LSDA, PBE, SCAN, and PBE-SIC, and $\Delta=-50\%$ for LSDA-SIC, constrained to values derived in the text.}
\label{fig:1}
\end{figure}

Using the self-consistent density for each functional and nearly-exact correlation energies\cite{Burke2016} for the rare-gas atoms Ne, Ar, Kr, and Xe, we have computed the relative errors of the various correlation energy functionals in Fig. \ref{fig:1}. We have then fitted them to the formula
\begin{equation}
\label{eq:12}
\frac{E_{c}^{approx}-E_{c}^{exact}}{E_{c}^{exact}} \times 100\% =\Delta+a\,x+b\,x^2+c\,x^3 \,\,,
\end{equation}
where $x=1/ln(Z)$. This formula was constructed by keeping just the first two terms in the expansion of Eq. (\ref{eq:5}), for both the exact and the approximate functionals, with the coefficients of the leading terms as derived in this section, making $\Delta=0\%$ for LSDA, PBE, SCAN, and PBE-SIC, but $\Delta=-50\%$ for LSDA-SIC. SCAN, which is already self-correlation-free, needs no SIC for correlation. From this, we constructed the large-$Z$ expansion of Eq. (\ref{eq:12}) for the relative error of each approximate functional. Enough terms in that expansion were kept to provide three parameters $a$, $b$, and $c$ to be least-squares fitted to the four considered atoms. (Four parameters would overfit the data.)

It is a long extrapolation from $Z=54$ (Xe) to $Z=\infty$. The data of Fig. \ref{fig:1} for each functional prefigure but do not pinpoint that functional's analytic large-$Z$ limit.

\begin{figure}
\centering
\includegraphics[width=1.0\linewidth]{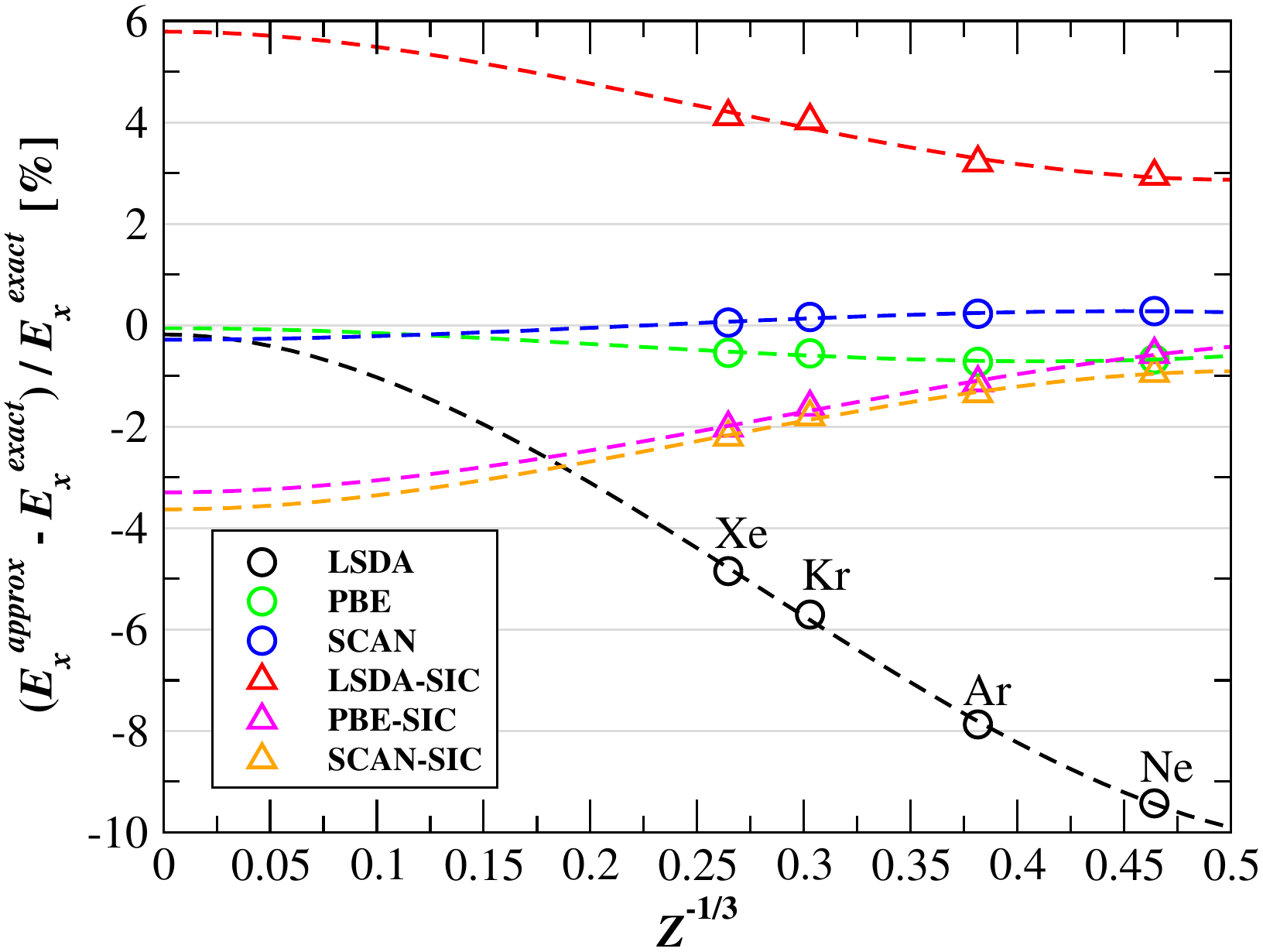}
\caption{Relative (\%) errors of six non-empirical functionals for the exchange energies of neutral atoms with atomic number $Z$. The points shown are for (from left to right) Xe ($Z=54$), Kr ($Z=36$), Ar ($Z=18$), and Ne ($Z=10$). The smooth curves are least-squares fits of the parameters $a$, $b$, and $c$ in the asymptotic formula of Eq. (\ref{eq:13}). The $Z^{-1/3}\rightarrow 0$ limit of the fit predicts (but does not constrain) the relative error of the exchange energy for a spin-unpolarized uniform electron gas.}
\label{fig:2}
\end{figure}

\begin{figure}[h!]
\centering
\includegraphics[width=1.0\linewidth]{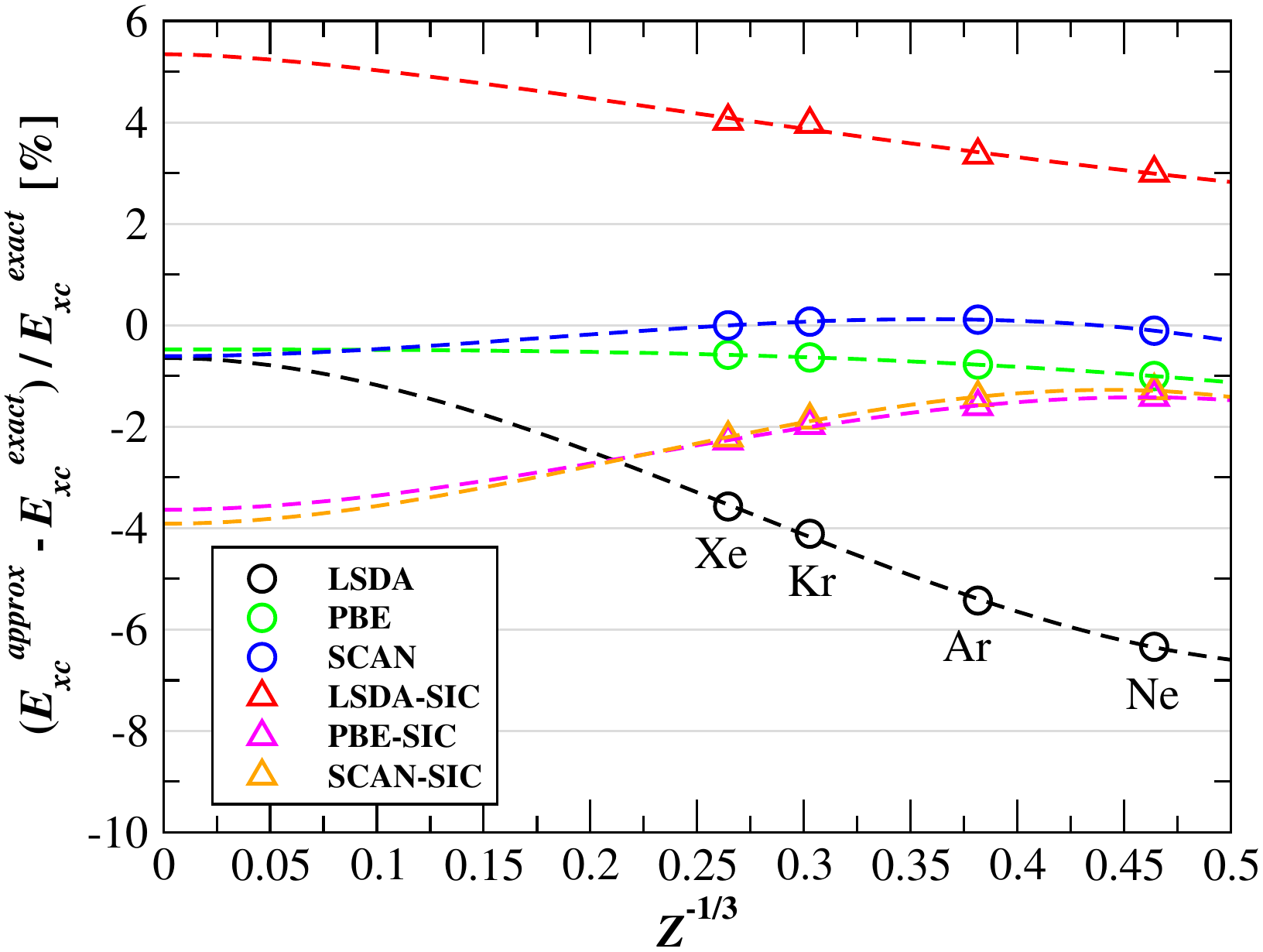}
\caption{Relative (\%) errors of six non-empirical functionals for the exchange-correlation energies of neutral atoms with atomic number $Z$. The points shown are for (from left to right) Xe ($Z=54$), Kr ($Z=36$), Ar ($Z=18$), and Ne ($Z=10$). The smooth curves are least-squares fits of the parameters $a$, $b$, and $c$ in the asymptotic formula of Eq. (\ref{eq:14}). The $Z^{-1/3}\rightarrow 0$ limit of the fit predicts (but does not constrain) the relative error of the exchange energy for a spin-unpolarized uniform electron gas.}
\label{fig:3}
\end{figure}

\section{\label{sec:3}Exchange and Exchange-Correlation energy: Extrapolation of the relative errors for N\lowercase{e}, A\lowercase{r}, K\lowercase{r}, and X\lowercase{e}}

Using the self-consistent density for each functional and Hartree-Fock exchange energies\cite{Becke1988} for the rare-gas atoms Ne, Ar, Kr, and Xe, we have plotted the relative errors of the various approximate exchange functionals in Fig. \ref{fig:2}. We have then fitted them to the formula
\begin{equation}
\label{eq:13}
\frac{E_{x}^{approx}-E_{x}^{exact}}{E_{x}^{exact}} \times 100\% = a + b \, x^2 + c \, x^3 \,\,,
\end{equation}
where $x=Z^{-1/3}$. This formula was constructed by keeping just the first three terms of Eq. (\ref{eq:4}) for both the exact and the approximate exchange energies (Appendix A). From this, we constructed the large-$Z$ expansion of Eq. (\ref{eq:13}), keeping just enough terms to provide three parameters $a$, $b$, and $c$ to be least-squares fitted to the four considered atoms.

Although we know that the parameter $a$ should be 0\% for LSDA, PBE, and SCAN, we have not constrained $a$ to this exact value. Values of $|a|$ less than 0.5\% emerge for all three approximations (Appendix B), suggesting that the considerably larger +5.8\% for LSDA-SIC, -3.3\% for PBE-SIC, and -3.6\% for SCAN-SIC can be trusted within $\pm0.5\%$. These large-$Z$ limits are clearly pre-figured in the data for Ne, Ar, Kr, and Xe.

The exchange-correlation energy is just the sum of the exchange and correlation energies, and is presented for completeness in Fig. \ref{fig:3}. The fitting formula is
\begin{align}
\label{eq:14}
\frac{E_{xc}^{approx}-E_{xc}^{exact}}{E_{xc}^{exact}} \times 100\% &= a + \left(\frac{3A_c}{A_x}\right) (a-\Delta) \, x^2 \, ln(x) \nonumber \\ & + \,b \, x^2 + c \, x^3 \,\,,
\end{align}
where $x=Z^{-1/3}$. $A_c$ and $A_x$ are exact values from Table \ref{tab:1}. $\Delta$ is 0\% for LSDA, PBE, SCAN, PBE-SIC, and SCAN-SIC, but -50\% for LSDA-SIC. Eq. (\ref{eq:14}) was constructed (Appendix A) almost as Eq. (\ref{eq:13}) was, but includes the exact $Z\,ln(Z)$ terms derived in section \ref{sec:2}. The $Z\rightarrow \infty$ or $x\rightarrow 0$ limit is dominated by exchange, and yields conclusions similar to those of Fig. \ref{fig:2}: LSDA-SIC makes a +5.3\% error in this limit, while PBE-SIC and SCAN-SIC respectively make -3.6\% and -3.9\% errors.

\section{\label{sec:4}Conclusion}
The Perdew-Zunger self-interaction correction (PZ SIC) to a semi-local density-functional approximation for the exchange-correlation energy imposes some additional exact constraints, such as exactness for all collections of separated one-electron densities and (nearly) linear variation of the total energy between adjacent integer electron numbers. But it loses other exact constraints or appropriate norms, such as exactness for all uniform densities and 0\% error for neutral atoms in the limit of large atomic number. This could explain PZ SIC's curious performance history of striking successes and equally striking failures, which are in a way the complements to those of the semi-local approximations.

The difficulty in constructing energy-minimizing but localized SIC orbitals for the uniform electron gas has led until recently to large uncertainties in the relative errors of PZ SIC for this important appropriate norm. We have found analytically that PZ LSDA-SIC correlation makes an error of -50\% for spin-unpolarized uniform densities in the high-density limit or for large-$Z$ atoms, and -100\% for fully-spin-polarized uniform densities in the high-density limit, independent of the details of the localized SIC orbitals. By extrapolating the relative error of the exchange energy from Ne, Ar, Kr and Xe, guided by the known asymptotics of the large-$Z$ limit, and employing real localized orbitals, we have estimated that for the uniform electron gas the LSDA-SIC exchange energy is about 5.5\% too negative, while the PBE-SIC and SCAN-SIC exchange energies are not negative enough by about 3.5\%. The estimated uncertainties due to extrapolation, based upon our results for the semi-local functionals without SIC, are less than $\pm$0.5\%.

The complementary strengths and weaknesses of PZ SIC and a semi-local functional like SCAN \cite{Sun2015} to which it is applied suggest that a widely-useful approximation might be found from a generalized self-interaction correction to SCAN (actually just to SCAN exchange, since SCAN is self-correlation-free) which preserves the best features of both, being accurate for real atoms, large-$Z$ atoms, and uniform densities. This possibility is being explored\cite{Perdew2019}.

We have seen that a knowledge of the asymptotic forms (\ref{eq:4}) and (\ref{eq:5}), and data from the sequence of real atoms Ne, Ar, Kr, and Xe, are sufficient to determine the large-$Z$ limits for the LSDA, PBE, and SCAN exchange energy functionals within $\pm0.5\%$. There is of course no guarantee that the SIC exchange functionals approach their large-$Z$ limits as rapidly as these semi-local functionals do. But we know that real atoms are also appropriate norms\cite{Jones2015,Cancio2018} for functionals beyond LSDA. Just correcting the large relative errors of SIC exchange for the real atoms Ne, Ar, Kr, and Xe (Fig. \ref{fig:2}) might improve the performance of SIC for the equilibrium properties of molecules and solids.


\section*{Acknowledgement}
BS, who co-designed the project and performed the calculations, was supported by the US Department of Energy, Office of Science, Basic Energy Sciences, under Award No. DE-SC0018331 as part of the Computational Chemical Sciences Program. JPP, who co-designed the project and wrote the first draft of the manuscript, was supported by the US National Science Foundation under Grant No. DMR-1607868.

\appendix

\section{Derivation of the fitting formulas, Eqs. (\ref{eq:14}) and (\ref{eq:13})}

We define the exact and the approximate exchange-correlation energy by considering the sum of the first three terms of the exact large-$Z$ asymptotic expansions for the exchange energy (Eq. (\ref{eq:4})) and the first two terms of the exact large-$Z$ asymptotic expansions for the correlation energy (Eq. (\ref{eq:5})).
\begin{align}
\label{eq:a1}
E_{xc}^{exact}&=-A_x^{exact} \, Z^{\frac{5}{3}} -A_c^{exact} \, Z \, ln(Z) + \nonumber \\
&B_{xc}^{exact} \, Z + C_x^{exact} \, Z^{\frac{2}{3}} \,\,,
\end{align}
\begin{align}
\label{eq:a2}
E_{xc}^{approx}&=-A_x^{approx} \, Z^{\frac{5}{3}} -A_c^{approx} \, Z \, ln(Z) + \nonumber \\
&B_{xc}^{approx} \, Z + C_x^{approx} \, Z^{\frac{2}{3}} \,\,.
\end{align}

The relative error is defined as
\begin{widetext}
\begin{equation}
\label{eq:a3}
\frac{E_{xc}^{approx}-E_{xc}^{exact}}{E_{xc}^{exact}}= \frac{\Delta A_x \, Z^{\frac{5}{3}} + \Delta A_c \,Z\,ln(Z) + \Delta B_{xc} \, Z + \Delta C_x \, Z^{\frac{2}{3}}}{-A_x^{exact} \, Z^{\frac{5}{3}} - A_c^{exact} \, Z \, ln(Z) + B_{xc}^{exact} \, Z + C_x^{exact} \, Z^{\frac{2}{3}}} \,,
\end{equation}
\end{widetext}
where $\Delta A_x=-(A_{x}^{approx}-A_{x}^{exact})$, $\Delta A_c=-(A_{c}^{approx}-A_{c}^{exact})$, $\Delta B_{xc}=B_{xc}^{approx}-B_{xc}^{exact}$, and $\Delta C_x=C_{x}^{approx}-C_{x}^{exact}$.

Now dividing numerator and denominator by the leading term $-A_x \, Z^{\frac{5}{3}}$ we get
\begin{widetext}
\begin{align}
\label{eq:a4}
\frac{E_{xc}^{approx}-E_{xc}^{exact}}{E_{xc}^{exact}}
&= \frac{-\left(\frac{\Delta A_x}{A_x}  + \frac{\Delta A_c}{A_x} \,Z^{-\frac{2}{3}}\,ln(Z) + \frac{\Delta B_{xc}}{A_x} \, Z^{-\frac{2}{3}} + \frac{\Delta C_x}{A_x} \, Z^{-1}\right)}{1 + \frac{A_c}{A_x} \, Z^{-\frac{2}{3}} \, ln(Z) - \frac{B_{xc}}{A_x} \, Z^{-\frac{2}{3}} - \frac{C_x}{A_x} \, Z^{-1}} \nonumber \\
&=-\left(\frac{\Delta A_x}{A_x}  + \frac{\Delta A_c}{A_x} \,Z^{-\frac{2}{3}}\,ln(Z) + \frac{\Delta B_{xc}}{A_x} \, Z^{-\frac{2}{3}} + \frac{\Delta C_x}{A_x} \, Z^{-1}\right) \nonumber \\
&+\frac{\Delta A_x}{A_x} \left(\frac{A_c}{A_x} \, Z^{-\frac{2}{3}} \, ln(Z) - \frac{B_{xc}}{A_x} \, Z^{-\frac{2}{3}} - \frac{C_x}{A_x} \, Z^{-1} \right) + \mathcal{O} \, Z^{-\frac{4}{3}} \, ln^2(Z)+ \cdots \,\,.
\end{align}
\end{widetext}

Neglecting the lower-order terms and using two free parameters, $b$ to replace $\left(-\frac{\Delta B_{xc}}{A_x}-\frac{B_{xc}\,\Delta A_x}{A_x^2}\right)$ and $c$ to replace $\left(-\frac{\Delta C_x}{A_x}-\frac{C_x\,\Delta A_x}{A_x^2}\right)$, we get:
\begin{widetext}
\begin{align}
\label{eq:a5}
\frac{E_{xc}^{approx}-E_{xc}^{exact}}{E_{xc}^{exact}}
=-\frac{\Delta A_x}{A_x} -\frac{A_c}{A_x}\left(-\frac{\Delta A_x}{A_x}+\frac{\Delta A_c}{A_c}\right) \, Z^{-\frac{2}{3}} \, ln(Z) + b \,  Z^{-\frac{2}{3}} + c \, Z^{-1} \,\,.
\end{align}
\end{widetext}

Now we consider another free parameter $a$ to replace $(-\frac{\Delta A_x}{A_x})$ and define $\Delta=-\frac{\Delta A_c}{A_c}$. Then we use $Z^{-\frac{1}{3}}=x$ and $ln(Z)=-3\,ln(x)$ to obtain
\begin{widetext}
\begin{align}
\label{eq:a6}
\frac{E_{xc}^{approx}-E_{xc}^{exact}}{E_{xc}^{exact}}=a +\left(\frac{3 A_c}{A_x}\right)\left(a-\Delta\right) \, x^2 \, ln(x) + b \, x^2 + c \,x^3 \,\,.
\end{align}
\end{widetext}

Here, $A_c$ and $A_x$ are exact values given in Table \ref{tab:1} of the main article. $\Delta$ is 0 for LSDA, PBE, SCAN, PBE-SIC, and SCAN-SIC. $\Delta$ is -50\% for LSDA-SIC since $A_c^{LSDA-SIC}=\frac{1}{2}\,A_c^{exact}$.

The above Eq. (\ref{eq:a6}) is the same as Eq. (\ref{eq:14}). When exchange-only terms are considered, the second term vanishes and the equation reduces to Eq. (\ref{eq:13}).

\section{Parameters of the fits to Eqs. (\ref{eq:13}) and (\ref{eq:14})}

The parameters ($a,b,c$) obtained from least-square fits to the asymptotic Eq. (\ref{eq:13}) are: LSDA ($-0.18, -95.90, 114.05$); PBE ($-0.06, -11.56, 18.73$); SCAN ($-0.28, 8.31, -12.29$); LSDA-SIC ($5.79,-34.86,46.38$); PBE-SIC ($-3.30,26.84,-30.71$); SCAN-SIC ($-3.63$, $32.12$, $-42.44$).

The parameters ($a,b,c$) obtained from least-square fits to the asymptotic Eq. (\ref{eq:14}) are: LSDA ($-0.64, -61.36, 74.87$); PBE ($-0.48, -0.54, -4.28$); SCAN ($-0.61, 16.68, -31.18$); LSDA-SIC ($5.34,5.05,-8.66$); PBE-SIC ($-3.64,29.83,-43.81$); SCAN-SIC ($-3.91$, $38.39$, $-58.30$).

We have also examined the effect of the choice of electron density on the fitted parameter $a$ which indicates the relative (\%) error in DFT exchange energy in the large-$Z$ limit (using Eq. (\ref{eq:13})). The parameter $a$ computed at self consistent density ($n_{scf}$) for LSDA, PBE, and SCAN functionals are $-0.18$, $-0.06$, and $-0.28$, respectively. In comparison, the parameter $a$ computed at the exact Hartree-Fock density\cite{Roetti1974} for LSDA, PBE, and SCAN functionals are $0.09$, $0.13$, and $-0.06$, respectively. The influence of the electron density is negligible in determining the parameter $a$ in Eq. (\ref{eq:13}).


\begin{thebibliography}{34}%
\makeatletter
\providecommand \@ifxundefined [1]{%
 \@ifx{#1\undefined}
}%
\providecommand \@ifnum [1]{%
 \ifnum #1\expandafter \@firstoftwo
 \else \expandafter \@secondoftwo
 \fi
}%
\providecommand \@ifx [1]{%
 \ifx #1\expandafter \@firstoftwo
 \else \expandafter \@secondoftwo
 \fi
}%
\providecommand \natexlab [1]{#1}%
\providecommand \enquote  [1]{``#1''}%
\providecommand \bibnamefont  [1]{#1}%
\providecommand \bibfnamefont [1]{#1}%
\providecommand \citenamefont [1]{#1}%
\providecommand \href@noop [0]{\@secondoftwo}%
\providecommand \href [0]{\begingroup \@sanitize@url \@href}%
\providecommand \@href[1]{\@@startlink{#1}\@@href}%
\providecommand \@@href[1]{\endgroup#1\@@endlink}%
\providecommand \@sanitize@url [0]{\catcode `\\12\catcode `\$12\catcode
  `\&12\catcode `\#12\catcode `\^12\catcode `\_12\catcode `\%12\relax}%
\providecommand \@@startlink[1]{}%
\providecommand \@@endlink[0]{}%
\providecommand \url  [0]{\begingroup\@sanitize@url \@url }%
\providecommand \@url [1]{\endgroup\@href {#1}{\urlprefix }}%
\providecommand \urlprefix  [0]{URL }%
\providecommand \Eprint [0]{\href }%
\providecommand \doibase [0]{http://dx.doi.org/}%
\providecommand \selectlanguage [0]{\@gobble}%
\providecommand \bibinfo  [0]{\@secondoftwo}%
\providecommand \bibfield  [0]{\@secondoftwo}%
\providecommand \translation [1]{[#1]}%
\providecommand \BibitemOpen [0]{}%
\providecommand \bibitemStop [0]{}%
\providecommand \bibitemNoStop [0]{.\EOS\space}%
\providecommand \EOS [0]{\spacefactor3000\relax}%
\providecommand \BibitemShut  [1]{\csname bibitem#1\endcsname}%
\let\auto@bib@innerbib\@empty
\bibitem [{\citenamefont {Kohn}\ and\ \citenamefont {Sham}(1965)}]{Kohn1965}%
  \BibitemOpen
  \bibfield  {author} {\bibinfo {author} {\bibfnamefont {W.}~\bibnamefont
  {Kohn}}\ and\ \bibinfo {author} {\bibfnamefont {L.~J.}\ \bibnamefont
  {Sham}},\ }\href {\doibase 10.1103/PhysRev.140.A1133} {\bibfield  {journal}
  {\bibinfo  {journal} {Phys. Rev.}\ }\textbf {\bibinfo {volume} {140}},\
  \bibinfo {pages} {A1133} (\bibinfo {year} {1965})}\BibitemShut {NoStop}%
\bibitem [{\citenamefont {Jones}(2015)}]{Jones2015}%
  \BibitemOpen
  \bibfield  {author} {\bibinfo {author} {\bibfnamefont {R.~O.}\ \bibnamefont
  {Jones}},\ }\href {\doibase 10.1103/RevModPhys.87.897} {\bibfield  {journal}
  {\bibinfo  {journal} {Rev. Mod. Phys.}\ }\textbf {\bibinfo {volume} {87}},\
  \bibinfo {pages} {897} (\bibinfo {year} {2015})}\BibitemShut {NoStop}%
\bibitem [{\citenamefont {Sun}\ \emph {et~al.}(2015)\citenamefont {Sun},
  \citenamefont {Ruzsinszky},\ and\ \citenamefont {Perdew}}]{Sun2015}%
  \BibitemOpen
  \bibfield  {author} {\bibinfo {author} {\bibfnamefont {J.}~\bibnamefont
  {Sun}}, \bibinfo {author} {\bibfnamefont {A.}~\bibnamefont {Ruzsinszky}}, \
  and\ \bibinfo {author} {\bibfnamefont {J.~P.~J.}\ \bibnamefont {Perdew}},\
  }\href {\doibase 10.1103/PhysRevLett.115.036402} {\bibfield  {journal}
  {\bibinfo  {journal} {Phys. Rev. Lett.}\ }\textbf {\bibinfo {volume} {115}},\
  \bibinfo {pages} {036402} (\bibinfo {year} {2015})}\BibitemShut {NoStop}%
\bibitem [{\citenamefont {Perdew}\ and\ \citenamefont
  {Wang}(1992)}]{Perdew1992}%
  \BibitemOpen
  \bibfield  {author} {\bibinfo {author} {\bibfnamefont {J.~P.}\ \bibnamefont
  {Perdew}}\ and\ \bibinfo {author} {\bibfnamefont {Y.}~\bibnamefont {Wang}},\
  }\href {\doibase 10.1103/PhysRevB.45.13244} {\bibfield  {journal} {\bibinfo
  {journal} {Phys. Rev. B}\ }\textbf {\bibinfo {volume} {45}},\ \bibinfo
  {pages} {13244} (\bibinfo {year} {1992})}\BibitemShut {NoStop}%
\bibitem [{\citenamefont {Perdew}\ and\ \citenamefont
  {Zunger}(1981)}]{Perdew1981}%
  \BibitemOpen
  \bibfield  {author} {\bibinfo {author} {\bibfnamefont {J.~P.}\ \bibnamefont
  {Perdew}}\ and\ \bibinfo {author} {\bibfnamefont {A.}~\bibnamefont
  {Zunger}},\ }\href {\doibase 10.1103/PhysRevB.23.5048} {\bibfield  {journal}
  {\bibinfo  {journal} {Phys. Rev. B}\ }\textbf {\bibinfo {volume} {23}},\
  \bibinfo {pages} {5048} (\bibinfo {year} {1981})}\BibitemShut {NoStop}%
\bibitem [{\citenamefont {Pederson}\ \emph {et~al.}(1984)\citenamefont
  {Pederson}, \citenamefont {Heaton},\ and\ \citenamefont
  {Lin}}]{Pederson1984}%
  \BibitemOpen
  \bibfield  {author} {\bibinfo {author} {\bibfnamefont {M.~R.}\ \bibnamefont
  {Pederson}}, \bibinfo {author} {\bibfnamefont {R.~A.}\ \bibnamefont
  {Heaton}}, \ and\ \bibinfo {author} {\bibfnamefont {C.~C.}\ \bibnamefont
  {Lin}},\ }\href {\doibase 10.1063/1.446959} {\bibfield  {journal} {\bibinfo
  {journal} {J. Chem. Phys.}\ }\textbf {\bibinfo {volume} {80}},\ \bibinfo
  {pages} {1972} (\bibinfo {year} {1984})}\BibitemShut {NoStop}%
\bibitem [{\citenamefont {Pederson}\ \emph {et~al.}(2014)\citenamefont
  {Pederson}, \citenamefont {Ruzsinszky},\ and\ \citenamefont
  {Perdew}}]{Pederson2014}%
  \BibitemOpen
  \bibfield  {author} {\bibinfo {author} {\bibfnamefont {M.~R.}\ \bibnamefont
  {Pederson}}, \bibinfo {author} {\bibfnamefont {A.}~\bibnamefont
  {Ruzsinszky}}, \ and\ \bibinfo {author} {\bibfnamefont {J.~P.}\ \bibnamefont
  {Perdew}},\ }\href {\doibase 10.1063/1.4869581} {\bibfield  {journal}
  {\bibinfo  {journal} {J. Chem. Phys.}\ }\textbf {\bibinfo {volume} {140}},\
  \bibinfo {pages} {121103} (\bibinfo {year} {2014})}\BibitemShut {NoStop}%
\bibitem [{\citenamefont {Perdew}\ \emph {et~al.}(1982)\citenamefont {Perdew},
  \citenamefont {Parr}, \citenamefont {Levy},\ and\ \citenamefont
  {Balduz}}]{Perdew1982}%
  \BibitemOpen
  \bibfield  {author} {\bibinfo {author} {\bibfnamefont {J.~P.}\ \bibnamefont
  {Perdew}}, \bibinfo {author} {\bibfnamefont {R.~G.}\ \bibnamefont {Parr}},
  \bibinfo {author} {\bibfnamefont {M.}~\bibnamefont {Levy}}, \ and\ \bibinfo
  {author} {\bibfnamefont {J.~L.}\ \bibnamefont {Balduz}},\ }\href {\doibase
  10.1103/PhysRevLett.49.1691} {\bibfield  {journal} {\bibinfo  {journal}
  {Phys. Rev. Lett.}\ }\textbf {\bibinfo {volume} {49}},\ \bibinfo {pages}
  {1691} (\bibinfo {year} {1982})}\BibitemShut {NoStop}%
\bibitem [{\citenamefont {Ruzsinszky}\ \emph {et~al.}(2006)\citenamefont
  {Ruzsinszky}, \citenamefont {Perdew}, \citenamefont {Csonka}, \citenamefont
  {Vydrov},\ and\ \citenamefont {Scuseria}}]{Ruzsinszky2006}%
  \BibitemOpen
  \bibfield  {author} {\bibinfo {author} {\bibfnamefont {A.}~\bibnamefont
  {Ruzsinszky}}, \bibinfo {author} {\bibfnamefont {J.~P.}\ \bibnamefont
  {Perdew}}, \bibinfo {author} {\bibfnamefont {G.~I.}\ \bibnamefont {Csonka}},
  \bibinfo {author} {\bibfnamefont {O.~A.}\ \bibnamefont {Vydrov}}, \ and\
  \bibinfo {author} {\bibfnamefont {G.~E.}\ \bibnamefont {Scuseria}},\ }\href
  {\doibase 10.1063/1.2387954} {\bibfield  {journal} {\bibinfo  {journal} {J.
  Chem. Phys.}\ }\textbf {\bibinfo {volume} {125}},\ \bibinfo {pages} {194112}
  (\bibinfo {year} {2006})}\BibitemShut {NoStop}%
\bibitem [{\citenamefont {Patchkovskii}\ and\ \citenamefont
  {Ziegler}(2002)}]{Patchkovskii2002}%
  \BibitemOpen
  \bibfield  {author} {\bibinfo {author} {\bibfnamefont {S.}~\bibnamefont
  {Patchkovskii}}\ and\ \bibinfo {author} {\bibfnamefont {T.}~\bibnamefont
  {Ziegler}},\ }\href {\doibase 10.1063/1.1468640} {\bibfield  {journal}
  {\bibinfo  {journal} {J. Chem. Phys.}\ }\textbf {\bibinfo {volume} {116}},\
  \bibinfo {pages} {7806} (\bibinfo {year} {2002})}\BibitemShut {NoStop}%
\bibitem [{\citenamefont {Perdew}\ \emph {et~al.}(1996)\citenamefont {Perdew},
  \citenamefont {Burke},\ and\ \citenamefont {Ernzerhof}}]{Perdew1996}%
  \BibitemOpen
  \bibfield  {author} {\bibinfo {author} {\bibfnamefont {J.~P.}\ \bibnamefont
  {Perdew}}, \bibinfo {author} {\bibfnamefont {K.}~\bibnamefont {Burke}}, \
  and\ \bibinfo {author} {\bibfnamefont {M.}~\bibnamefont {Ernzerhof}},\ }\href
  {\doibase 10.1103/PhysRevLett.77.3865} {\bibfield  {journal} {\bibinfo
  {journal} {Phys. Rev. Lett.}\ }\textbf {\bibinfo {volume} {77}},\ \bibinfo
  {pages} {3865} (\bibinfo {year} {1996})}\BibitemShut {NoStop}%
\bibitem [{\citenamefont {Vydrov}\ and\ \citenamefont
  {Scuseria}(2004)}]{Vydrov2004}%
  \BibitemOpen
  \bibfield  {author} {\bibinfo {author} {\bibfnamefont {O.~A.}\ \bibnamefont
  {Vydrov}}\ and\ \bibinfo {author} {\bibfnamefont {G.~E.}\ \bibnamefont
  {Scuseria}},\ }\href {\doibase 10.1063/1.1794633} {\bibfield  {journal}
  {\bibinfo  {journal} {J. Chem. Phys.}\ }\textbf {\bibinfo {volume} {121}},\
  \bibinfo {pages} {8187} (\bibinfo {year} {2004})}\BibitemShut {NoStop}%
\bibitem [{\citenamefont {Vydrov}\ \emph {et~al.}(2006)\citenamefont {Vydrov},
  \citenamefont {Scuseria}, \citenamefont {Perdew}, \citenamefont
  {Ruzsinszky},\ and\ \citenamefont {Csonka}}]{Vydrov2006}%
  \BibitemOpen
  \bibfield  {author} {\bibinfo {author} {\bibfnamefont {O.~A.}\ \bibnamefont
  {Vydrov}}, \bibinfo {author} {\bibfnamefont {G.~E.}\ \bibnamefont
  {Scuseria}}, \bibinfo {author} {\bibfnamefont {J.~P.}\ \bibnamefont
  {Perdew}}, \bibinfo {author} {\bibfnamefont {A.}~\bibnamefont {Ruzsinszky}},
  \ and\ \bibinfo {author} {\bibfnamefont {G.~I.}\ \bibnamefont {Csonka}},\
  }\href {\doibase 10.1063/1.2176608} {\bibfield  {journal} {\bibinfo
  {journal} {J. Chem. Phys.}\ }\textbf {\bibinfo {volume} {124}},\ \bibinfo
  {pages} {094108} (\bibinfo {year} {2006})}\BibitemShut {NoStop}%
\bibitem [{\citenamefont {Shahi}\ \emph {et~al.}()\citenamefont {Shahi},
  \citenamefont {Bhattarai}, \citenamefont {Wagle}, \citenamefont {Santra},
  \citenamefont {Schwalbe}, \citenamefont {Hahn}, \citenamefont {Kortus},
  \citenamefont {Jackson}, \citenamefont {Peralta}, \citenamefont {Trepte},
  \citenamefont {Lehtola}, \citenamefont {Nepal}, \citenamefont {Myneni},
  \citenamefont {Neupane}, \citenamefont {Adhikari}, \citenamefont
  {Ruzsinszky}, \citenamefont {Yamamoto}, \citenamefont {Baruah}, \citenamefont
  {Zope},\ and\ \citenamefont {Perdew}}]{Shahi2019}%
  \BibitemOpen
  \bibfield  {author} {\bibinfo {author} {\bibfnamefont {C.}~\bibnamefont
  {Shahi}}, \bibinfo {author} {\bibfnamefont {P.}~\bibnamefont {Bhattarai}},
  \bibinfo {author} {\bibfnamefont {K.}~\bibnamefont {Wagle}}, \bibinfo
  {author} {\bibfnamefont {B.}~\bibnamefont {Santra}}, \bibinfo {author}
  {\bibfnamefont {S.}~\bibnamefont {Schwalbe}}, \bibinfo {author}
  {\bibfnamefont {T.}~\bibnamefont {Hahn}}, \bibinfo {author} {\bibfnamefont
  {J.}~\bibnamefont {Kortus}}, \bibinfo {author} {\bibfnamefont {K.~A.}\
  \bibnamefont {Jackson}}, \bibinfo {author} {\bibfnamefont {J.~E.}\
  \bibnamefont {Peralta}}, \bibinfo {author} {\bibfnamefont {K.}~\bibnamefont
  {Trepte}}, \bibinfo {author} {\bibfnamefont {S.}~\bibnamefont {Lehtola}},
  \bibinfo {author} {\bibfnamefont {N.~K.}\ \bibnamefont {Nepal}}, \bibinfo
  {author} {\bibfnamefont {H.}~\bibnamefont {Myneni}}, \bibinfo {author}
  {\bibfnamefont {B.}~\bibnamefont {Neupane}}, \bibinfo {author} {\bibfnamefont
  {S.}~\bibnamefont {Adhikari}}, \bibinfo {author} {\bibfnamefont
  {A.}~\bibnamefont {Ruzsinszky}}, \bibinfo {author} {\bibfnamefont
  {Y.}~\bibnamefont {Yamamoto}}, \bibinfo {author} {\bibfnamefont
  {T.}~\bibnamefont {Baruah}}, \bibinfo {author} {\bibfnamefont {R.~R.}\
  \bibnamefont {Zope}}, \ and\ \bibinfo {author} {\bibfnamefont {J.~P.}\
  \bibnamefont {Perdew}},\ }\href {https://arxiv.org/abs/1903.00611} {\bibinfo
  {journal} {arXiv:1903.00611}\ }\BibitemShut {NoStop}%
\bibitem [{\citenamefont {Kl{\"{u}}pfel}\ \emph {et~al.}(2011)\citenamefont
  {Kl{\"{u}}pfel}, \citenamefont {Kl{\"{u}}pfel},\ and\ \citenamefont
  {J{\'{o}}nsson}}]{Klupfel2011}%
  \BibitemOpen
  \bibfield  {author} {\bibinfo {author} {\bibfnamefont {S.}~\bibnamefont
  {Kl{\"{u}}pfel}}, \bibinfo {author} {\bibfnamefont {P.}~\bibnamefont
  {Kl{\"{u}}pfel}}, \ and\ \bibinfo {author} {\bibfnamefont {H.}~\bibnamefont
  {J{\'{o}}nsson}},\ }\href {\doibase 10.1103/PhysRevA.84.050501} {\bibfield
  {journal} {\bibinfo  {journal} {Phys. Rev. A}\ }\textbf {\bibinfo {volume}
  {84}},\ \bibinfo {pages} {050501} (\bibinfo {year} {2011})}\BibitemShut
  {NoStop}%
\bibitem [{\citenamefont {Perdew}\ \emph {et~al.}(2006)\citenamefont {Perdew},
  \citenamefont {Constantin}, \citenamefont {Sagvolden},\ and\ \citenamefont
  {Burke}}]{Perdew2006}%
  \BibitemOpen
  \bibfield  {author} {\bibinfo {author} {\bibfnamefont {J.~P.}\ \bibnamefont
  {Perdew}}, \bibinfo {author} {\bibfnamefont {L.~A.}\ \bibnamefont
  {Constantin}}, \bibinfo {author} {\bibfnamefont {E.}~\bibnamefont
  {Sagvolden}}, \ and\ \bibinfo {author} {\bibfnamefont {K.}~\bibnamefont
  {Burke}},\ }\href {\doibase 10.1103/PhysRevLett.97.223002} {\bibfield
  {journal} {\bibinfo  {journal} {Phys. Rev. Lett.}\ }\textbf {\bibinfo
  {volume} {97}},\ \bibinfo {pages} {223002} (\bibinfo {year}
  {2006})}\BibitemShut {NoStop}%
\bibitem [{\citenamefont {Burke}\ \emph {et~al.}(2016)\citenamefont {Burke},
  \citenamefont {Cancio}, \citenamefont {Gould},\ and\ \citenamefont
  {Pittalis}}]{Burke2016}%
  \BibitemOpen
  \bibfield  {author} {\bibinfo {author} {\bibfnamefont {K.}~\bibnamefont
  {Burke}}, \bibinfo {author} {\bibfnamefont {A.}~\bibnamefont {Cancio}},
  \bibinfo {author} {\bibfnamefont {T.}~\bibnamefont {Gould}}, \ and\ \bibinfo
  {author} {\bibfnamefont {S.}~\bibnamefont {Pittalis}},\ }\href {\doibase
  10.1063/1.4959126} {\bibfield  {journal} {\bibinfo  {journal} {J. Chem.
  Phys.}\ }\textbf {\bibinfo {volume} {145}},\ \bibinfo {pages} {054112}
  (\bibinfo {year} {2016})}\BibitemShut {NoStop}%
\bibitem [{\citenamefont {Schwinger}(1981)}]{Schwinger1981}%
  \BibitemOpen
  \bibfield  {author} {\bibinfo {author} {\bibfnamefont {J.}~\bibnamefont
  {Schwinger}},\ }\href {\doibase 10.1103/PhysRevA.24.2353} {\bibfield
  {journal} {\bibinfo  {journal} {Phys. Rev. A}\ }\textbf {\bibinfo {volume}
  {24}},\ \bibinfo {pages} {2353} (\bibinfo {year} {1981})}\BibitemShut
  {NoStop}%
\bibitem [{\citenamefont {Elliott}\ and\ \citenamefont
  {Burke}(2009)}]{Elliott2009}%
  \BibitemOpen
  \bibfield  {author} {\bibinfo {author} {\bibfnamefont {P.}~\bibnamefont
  {Elliott}}\ and\ \bibinfo {author} {\bibfnamefont {K.}~\bibnamefont
  {Burke}},\ }\href {\doibase 10.1139/V09-095} {\bibfield  {journal} {\bibinfo
  {journal} {Can. J. Chem.}\ }\textbf {\bibinfo {volume} {87}},\ \bibinfo
  {pages} {1485} (\bibinfo {year} {2009})}\BibitemShut {NoStop}%
\bibitem [{\citenamefont {Cancio}\ \emph {et~al.}(2018)\citenamefont {Cancio},
  \citenamefont {Chen}, \citenamefont {Krull},\ and\ \citenamefont
  {Burke}}]{Cancio2018}%
  \BibitemOpen
  \bibfield  {author} {\bibinfo {author} {\bibfnamefont {A.}~\bibnamefont
  {Cancio}}, \bibinfo {author} {\bibfnamefont {G.~P.}\ \bibnamefont {Chen}},
  \bibinfo {author} {\bibfnamefont {B.~T.}\ \bibnamefont {Krull}}, \ and\
  \bibinfo {author} {\bibfnamefont {K.}~\bibnamefont {Burke}},\ }\href
  {\doibase 10.1063/1.5021597} {\bibfield  {journal} {\bibinfo  {journal} {J.
  Chem. Phys.}\ }\textbf {\bibinfo {volume} {149}},\ \bibinfo {pages} {084116}
  (\bibinfo {year} {2018})}\BibitemShut {NoStop}%
\bibitem [{\citenamefont {Norman}(1983)}]{Norman1983}%
  \BibitemOpen
  \bibfield  {author} {\bibinfo {author} {\bibfnamefont {M.~R.}\ \bibnamefont
  {Norman}},\ }\href {\doibase 10.1103/PhysRevB.28.3585} {\bibfield  {journal}
  {\bibinfo  {journal} {Phys. Rev. B}\ }\textbf {\bibinfo {volume} {28}},\
  \bibinfo {pages} {3585} (\bibinfo {year} {1983})}\BibitemShut {NoStop}%
\bibitem [{\citenamefont {Pederson}\ \emph {et~al.}(1989)\citenamefont
  {Pederson}, \citenamefont {Heaton},\ and\ \citenamefont
  {Harrison}}]{Pederson1989}%
  \BibitemOpen
  \bibfield  {author} {\bibinfo {author} {\bibfnamefont {M.~R.}\ \bibnamefont
  {Pederson}}, \bibinfo {author} {\bibfnamefont {R.~A.}\ \bibnamefont
  {Heaton}}, \ and\ \bibinfo {author} {\bibfnamefont {J.~G.}\ \bibnamefont
  {Harrison}},\ }\href {\doibase 10.1103/PhysRevB.39.1581} {\bibfield
  {journal} {\bibinfo  {journal} {Phys. Rev. B}\ }\textbf {\bibinfo {volume}
  {39}},\ \bibinfo {pages} {1581} (\bibinfo {year} {1989})}\BibitemShut
  {NoStop}%
\bibitem [{\citenamefont {G{\"{o}}rling}\ and\ \citenamefont
  {Ernzerhof}(1995)}]{Gorling1995}%
  \BibitemOpen
  \bibfield  {author} {\bibinfo {author} {\bibfnamefont {A.}~\bibnamefont
  {G{\"{o}}rling}}\ and\ \bibinfo {author} {\bibfnamefont {M.}~\bibnamefont
  {Ernzerhof}},\ }\href {\doibase 10.1103/PhysRevA.51.4501} {\bibfield
  {journal} {\bibinfo  {journal} {Phys. Rev. A}\ }\textbf {\bibinfo {volume}
  {51}},\ \bibinfo {pages} {4501} (\bibinfo {year} {1995})}\BibitemShut
  {NoStop}%
\bibitem [{\citenamefont {Levy}\ and\ \citenamefont {Perdew}(1985)}]{Levy1985}%
  \BibitemOpen
  \bibfield  {author} {\bibinfo {author} {\bibfnamefont {M.}~\bibnamefont
  {Levy}}\ and\ \bibinfo {author} {\bibfnamefont {J.~P.}\ \bibnamefont
  {Perdew}},\ }\href {\doibase 10.1103/PhysRevA.32.2010} {\bibfield  {journal}
  {\bibinfo  {journal} {Phys. Rev. A}\ }\textbf {\bibinfo {volume} {32}},\
  \bibinfo {pages} {2010} (\bibinfo {year} {1985})}\BibitemShut {NoStop}%
\bibitem [{\citenamefont {Becke}(1988)}]{Becke1988}%
  \BibitemOpen
  \bibfield  {author} {\bibinfo {author} {\bibfnamefont {A.~D.}\ \bibnamefont
  {Becke}},\ }\href {\doibase 10.1103/PhysRevA.38.3098} {\bibfield  {journal}
  {\bibinfo  {journal} {Phys. Rev. A}\ }\textbf {\bibinfo {volume} {38}},\
  \bibinfo {pages} {3098} (\bibinfo {year} {1988})}\BibitemShut {NoStop}%
\bibitem [{Per()}]{Perdew2019}%
  \BibitemOpen
  \href@noop {} {}\bibinfo {note} {J. P. Perdew, B. Santra, and T. Hahn, Work
  in progress}\BibitemShut {NoStop}%
\bibitem [{flo()}]{flosic0.1}%
  \BibitemOpen
  \href@noop {} {}\bibinfo {note} {FLOSIC 0.1, developed by R. R. Zope, T.
  Baruah, J. E. Peralta, and K. A. Jackson}\BibitemShut {NoStop}%
\bibitem [{\citenamefont {Porezag}\ and\ \citenamefont
  {Pederson}(1999)}]{Porezag1999}%
  \BibitemOpen
  \bibfield  {author} {\bibinfo {author} {\bibfnamefont {D.}~\bibnamefont
  {Porezag}}\ and\ \bibinfo {author} {\bibfnamefont {M.~R.}\ \bibnamefont
  {Pederson}},\ }\href {\doibase 10.1103/PhysRevA.60.2840} {\bibfield
  {journal} {\bibinfo  {journal} {Phys. Rev. A}\ }\textbf {\bibinfo {volume}
  {60}},\ \bibinfo {pages} {2840} (\bibinfo {year} {1999})}\BibitemShut
  {NoStop}%
\bibitem [{\citenamefont {Pederson}\ and\ \citenamefont
  {Jackson}(1990)}]{Pederson1990}%
  \BibitemOpen
  \bibfield  {author} {\bibinfo {author} {\bibfnamefont {M.~R.}\ \bibnamefont
  {Pederson}}\ and\ \bibinfo {author} {\bibfnamefont {K.~A.}\ \bibnamefont
  {Jackson}},\ }\href {\doibase 10.1103/PhysRevB.41.7453} {\bibfield  {journal}
  {\bibinfo  {journal} {Phys. Rev. B}\ }\textbf {\bibinfo {volume} {41}},\
  \bibinfo {pages} {7453} (\bibinfo {year} {1990})}\BibitemShut {NoStop}%
\bibitem [{\citenamefont {Pederson}(2015)}]{Pederson2015}%
  \BibitemOpen
  \bibfield  {author} {\bibinfo {author} {\bibfnamefont {M.~R.}\ \bibnamefont
  {Pederson}},\ }\href {\doibase 10.1063/1.4907592} {\bibfield  {journal}
  {\bibinfo  {journal} {J. Chem. Phys.}\ }\textbf {\bibinfo {volume} {142}},\
  \bibinfo {pages} {064112} (\bibinfo {year} {2015})}\BibitemShut {NoStop}%
\bibitem [{\citenamefont {Yang}\ \emph {et~al.}(2017)\citenamefont {Yang},
  \citenamefont {Pederson},\ and\ \citenamefont {Perdew}}]{Yang2017}%
  \BibitemOpen
  \bibfield  {author} {\bibinfo {author} {\bibfnamefont {Z.-H.}\ \bibnamefont
  {Yang}}, \bibinfo {author} {\bibfnamefont {M.~R.}\ \bibnamefont {Pederson}},
  \ and\ \bibinfo {author} {\bibfnamefont {J.~P.}\ \bibnamefont {Perdew}},\
  }\href {\doibase 10.1103/PhysRevA.95.052505} {\bibfield  {journal} {\bibinfo
  {journal} {Phys. Rev. A}\ }\textbf {\bibinfo {volume} {95}},\ \bibinfo
  {pages} {052505} (\bibinfo {year} {2017})}\BibitemShut {NoStop}%
\bibitem [{\citenamefont {Gell-Mann}\ and\ \citenamefont
  {Brueckner}(1957)}]{Gell-Mann1957}%
  \BibitemOpen
  \bibfield  {author} {\bibinfo {author} {\bibfnamefont {M.}~\bibnamefont
  {Gell-Mann}}\ and\ \bibinfo {author} {\bibfnamefont {K.~A.}\ \bibnamefont
  {Brueckner}},\ }\href {\doibase 10.1103/PhysRev.106.364} {\bibfield
  {journal} {\bibinfo  {journal} {Phys. Rev.}\ }\textbf {\bibinfo {volume}
  {106}},\ \bibinfo {pages} {364} (\bibinfo {year} {1957})}\BibitemShut
  {NoStop}%
\bibitem [{\citenamefont {Misawa}(1965)}]{Misawa1965}%
  \BibitemOpen
  \bibfield  {author} {\bibinfo {author} {\bibfnamefont {S.}~\bibnamefont
  {Misawa}},\ }\href {\doibase 10.1103/PhysRev.140.A1645} {\bibfield  {journal}
  {\bibinfo  {journal} {Phys. Rev.}\ }\textbf {\bibinfo {volume} {140}},\
  \bibinfo {pages} {A1645} (\bibinfo {year} {1965})}\BibitemShut {NoStop}%
\bibitem [{\citenamefont {Roetti}\ and\ \citenamefont
  {Clementi}(1974)}]{Roetti1974}%
  \BibitemOpen
  \bibfield  {author} {\bibinfo {author} {\bibfnamefont {C.}~\bibnamefont
  {Roetti}}\ and\ \bibinfo {author} {\bibfnamefont {E.}~\bibnamefont
  {Clementi}},\ }\href {\doibase 10.1063/1.1681531} {\bibfield  {journal}
  {\bibinfo  {journal} {J. Chem. Phys.}\ }\textbf {\bibinfo {volume} {60}},\
  \bibinfo {pages} {3342} (\bibinfo {year} {1974})}\BibitemShut {NoStop}%
\end{thebibliography}
%

\end{document}